\documentclass[aps,prl,twocolumn,groupedaddress]{revtex4}

\usepackage[dvips]{graphicx}

\begin{document}

\title{Unconventional Upper- and Lower-Critical Fields and Normal-State Magnetic Susceptibility of the Novel Superconducting Compound Na$_{0.35}$CoO$_{2}$$\cdot$1.3H$_{2}$O}

\author{Hiroya Sakurai, 
Kazunori Takada$^{1}$, Shunsuke Yoshii$^{2}$, Takayoshi Sasaki$^{1}$, Koichi Kindo$^{2}$, and Eiji Takayama-Muromachi}
\email[$e$-mail: ]{SAKURAI.Hiroya@nims.go.jp}

\affiliation{Superconducting Materials Center, National Institute for Materials Science (SMC/NIMS), 1-1 Namiki, Tsukuba, Ibaraki 305-0044, Japan\\
$^{1}$ Advanced Materials Laboratory, National Institute for Materials Science (AML/NIMS), 1-1 Namiki, Tsukuba, Ibaraki 305-0044, Japan\\
/CREST, Japan Science and Technology Corporation\\
$^{2}$ KYOKUGEN, Osaka University, Toyonaka, Osaka 560-8531, Japan}

\date{\today}

\begin{abstract}
Magnetic properties of the novel layered superconductor, Na$_{0.35}$CoO$_{2}$$\cdot$1.3H$_{2}$O have been investigated. From the temperature dependence and field dependence of the magnetization, the superconducting transition temperature, as well as upper- and lower-critical fields have been estimated to be $T_{C}=4.6$ K, $H_{C2}(0)=61.0$ T and $H_{C1}(0)=28.1$ Oe. These values give quite unusual phenomenological parameters, \textit{i.e.}, coherent length, penetration depth and Ginzburg-Landau parameter of  $\xi=2.32$ nm, $\lambda=5.68\times 10^{3}$ \AA\ and $\kappa\equiv\lambda/\xi=244$,  suggesting an unconventional nature of superconductivity. Normal-state magnetic susceptibility shows an upturn below 130 K, which is confirmed to be inherent by high-field magnetization data. The upturn may have relevance to the mechanism of the superconductivity.
\end{abstract}


\keywords{Na$_{0.35}$CoO$_{2}$$\cdot$1.3H$_{2}$O, Co, triangular lattice, RVB, superconductor, magnetic susceptibility, upper-critical field, lower-critical field, high field magnetization}

\maketitle

Very recently, a novel Co oxide, Na$_{0.35}$CoO$_{2}$$\cdot$1.3H$_{2}$O, was found to be a superconductor with $T_{C}\simeq$5 K\cite{Nature}. This compound is the first to be discovered superconducting Co oxide and moreover, it has been claimed that there is a marked resemblance between this compound and high $T_{C}$ cuprates\cite{Nature,NatureMat}. The first similarity is that the compound has a layered structure composed of two-dimensional (2D) CoO$_{2}$ planes, separated by a thick insulating layer of Na ions and water molecules. As is widely known, high $T_{C}$ cuprates have also layered structures with 2D CuO$_{2}$ planes which play an essential role in the high $T_{C}$ superconductivity. The second similarity is related to valences and spin states of the transition metals.  The Cu ions of the high $T_{C}$ cuprates are of mixed-valence states by hole doping; Cu$^{2+}$ (S=1/2) ions are partially oxidized to Cu$^{3+}$ (S=0) ions. In a similar way, the Co ion in the Co oxide is either Co$^{4+}$ (d$^{5}$) or Co$^{3+}$ (d$^{6}$) with a spin state of S=1/2 or S=0, respectively, assuming low-spin configurations.

It is also worth noting that there is an obvious difference between the two systems. In a high $T_{C}$ cuprate, the Cu atoms form a square lattice, whereas the Co atoms form a triangular lattice in the present material. This geometric difference appears to be quite important because it is related to the magnetic ground states of the mother compounds. In the nondoped CuO$_{2}$ planes with Cu$^{2+}$ (S=1/2) ions, antiferromagnetic long-range order is the ground state, but such a simple picture is not applicable for the hypothetical nondoped CoO$_{2}$ plane with Co$^{4+}$ (S=1/2) ions because of geometric frustration expected for the triangular lattice. Thus, it is not clear whether the ground state of the CoO$_{2}$ plane is an antiferromagnetically ordered state or a resonating-valence-bonds (RVB) state\cite{RVB}. An additional difference is that the holes of a high $T_{C}$ cuprate exist in the \textit{e$_{g}$} orbitals, while the electrons of the Co oxide exist in the \textit{t$_{2g}$} orbitals.

Studies of the present Co oxide have just begun, and its fundamental superconducting and normal-state properties still need to be investigated. In particular, magnetic properties seem to be of primary importance.  Such investigation may lead to understanding of superconductivity not only of the Co oxide but also of high $T_{C}$ cuprates. Here, we report upper- and lower-critical fields and normal-state magnetic susceptibilities up to 55 T for the superconducting Co oxide and they are discussed in comparison with those of conventional superconductors and high $T_{C}$ cuprates. 

A powder sample was synthesized as described in the previous report\cite{Nature}. The sample was shown by powder X-ray diffraction to be single phase with Na$_{0.35}$CoO$_{2}$$\cdot$1.3H$_{2}$O.
The magnetic data below 7 T were collected with a SQUID magnetometer (Quantum Design MPMS-XL7). All the measurements were done under zero-field-cooling condition.
The high-field magnetization measurements were performed by using a pulse magnet at KYOKUGEN in Osaka University.

\begin{figure}
\includegraphics[width=7cm,keepaspectratio]{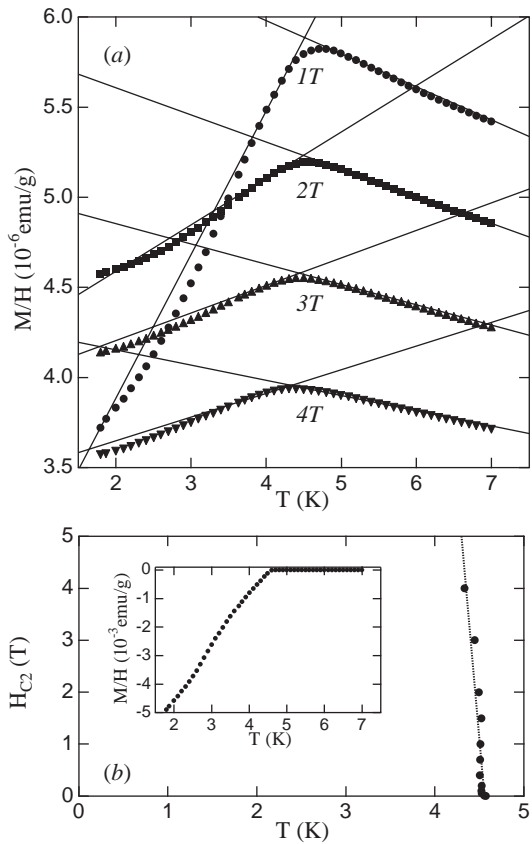}
\caption{
(a) Temperature dependence of the magnetization under various fields. Each curve except that under 1 T is offset for clarity. (b) Field dependence of $T_{C}$. The dotted line is the result when fitted by a linear function. The inset shows $M/H$ data at 20 Oe.
}
\label{Hc2}
\end{figure}
Typical magnetization/magnetic field ($M/H$) data are shown as a function of temperature in Fig. \ref{Hc2}($a$). For $H\geq 7$ kOe, magnetic susceptibility was not negative down to 1.8 K but the superconducting transition was observed as a downturn of $M/H$. The transition temperature ($T_{C}$) could be determined by the simple straight line fit as shown in Fig. \ref{Hc2}($a$), for given $H$ of 0.02, 0.1, 0.5, 1, 2, 4, 7, 10, 15, 20, 30, and 40 kOe. For $H\geq 50$ kOe, however, the superconducting transition became quite broad and $T_{C}$ could not be determined without seriously large uncertainty.  This procedure gives an upper-critical field ($H_{C2}$) versus temperature relation of $H_{C2}(T)=88.0-19.3T$, as shown in Fig. \ref{Hc2}($b$). From this equation, $T_{C}$ at $H=0$ and $dH_{C2}/dT\mid_{T_{C}}$ are calculated to be 4.56 K and 19.3 T/K, respectively. The value of $T_{C}$ obtained at $H=0$ agrees well with the onset $T_{C}$ determined from the susceptibility data at $H=20$ Oe (see the insert of Fig. \ref{Hc2}($b$)) but slightly lower than that in the previous report\cite{Nature}. The initial slope, $dH_{C2}/dT\mid_{T_{C}}$, is much larger than the $\sim$7.5 T/K of a molybdenum chalcogenide which had been known to have the highest initial slope before the discovery of high $T_{C}$ cuprates\cite{MoComp}.

According to the Werthamer-Helfand-Hohenberg (WHH) formula\cite{estHc2},
\begin{equation}
H_{C2}(0)=0.693(\frac{dH_{C2}}{dT}\mid_{T_{C}})T_{C},
\label{eqHc2}
\end{equation}
$H_{C2}(0)$ is calculated to be 61.0 T. In high $T_{C}$ cuprates, $H_{C2}$ does not always have a strictly defined meaning. This may also be the case for the present system. Corresponding to the high $H_{C2}$ value, a quite small coherent length, $\xi=2.32$ nm, is obtained according to the formula, $H_{C2}=\Phi_{0}/2\pi \xi^{2}$ ($\Phi_{0}$: fluxoid quantum). This length is comparable to those of high $T_{C}$ cuprates but far smaller than those of conventional superconductors. The present compound has highly 2D structure with a hexagonal lattice and strong anisotropy is expected for the superconducting properties.  Since $H_{C2}^{H//c}\ll H_{C2}^{H//ab}$ by analogy of high $T_{C}$ cuprates, $H_{C2}$ determined above reflects $H_{C2}^{H//ab}(=\Phi _{0}/2\pi \xi_{ab}\xi_{c}$) and $\xi=2.32$ nm is an average of  $\sqrt{\xi_{ab}\xi_{c}}$ ($\xi_{i}$: the coherent length of $i$-direction). For further discussion of the anisotropy, we need single crystal data, though single crystal growth seems quite hard for the present system.

\begin{figure}
\includegraphics[width=7cm,keepaspectratio]{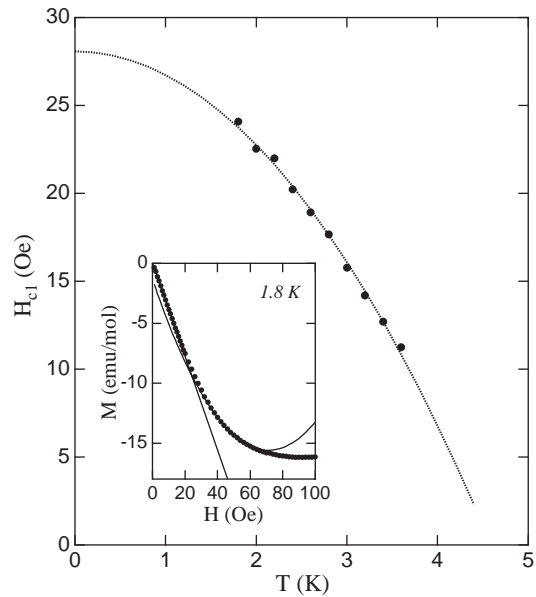}
\caption{
Temperature dependences of $H_{C1}$. The dotted line is the result when fitted by Eq. \ref{eqHc1}. The inset shows the magnetization curve at 1.8 K. The solid lines are the linear function and the parabola for the estimation of $H_{C1}$ (see the text).
}
\label{Hc1}
\end{figure}
The temperature dependence of the lower-critical field $H_{C1}$ is shown in Fig. \ref{Hc1}. The $H_{C1}$ values were determined from $M$-$H$ curves below 100 Oe at various temperatures. The magnetization decreases as $M=\chi H$ with increasing applied field below $H_{C1}$ and starts to deviate from the straight line at $H_{C1}$. According to the Bean's critical-state model\cite{Bean}, the deviation, $\Delta M$, from the linear function may be calculated as $\sqrt{\Delta M}\propto -H_{C1}+H$\cite{estHc1}. However, $\sqrt{\Delta M}$ of our data did not obey well this linear relation. This might be due to the shape effect which is not taken into account in the model. The simple straight line fit used in the $H_{C2}$ determination was not appropriate to estimate $H_{C1}$ because the result depended largely on the range of fitting for the data above $H_{C1}$. Instead, we represented the $M$-$H$ curve above $H_{C1}$ by a quadratic function and $H_{C1}(T)$ was determined from the intersection point of the quadratic curve and the straight line which was obtained for the low field data. 
The $H_{C1}(T)$ values obtained were fitted by the function:
\begin{equation}
H_{C1}(T)=H_{C1}(0)[1-(T/T_{C})^2],
\label{eqHc1}
\end{equation}
where $H_{C1}(0)$ and $T_{C}$ were fitting parameters, resulting in $H_{C1}(0)=28.1$ Oe and $T_{C}=4.59$ K. The $T_{C}$ thus calculated agrees very well with that determined from $H_{C2}$. 

From $H_{C1}(0)$ and $\xi$, the penetration depth can be calculated to be $\lambda=5.68\times 10^{3}$ \AA\ by the formula,
\begin{equation}
H_{C1}=\frac{\Phi _{0}}{4\pi\lambda^{2}}\ln(\frac{\lambda}{\xi}).
\end{equation}
The Ginzburg-Landau (GL) parameter, $\kappa\equiv\lambda/\xi=244$, is much larger than the parameters of conventional superconductors (it is even larger than those of high $T_{C}$ cuprates), suggesting that the present compound belongs to an extreme type II family.  Since $H_{C1}^{H//c}>H_{C1}^{H//ab}$ is expected by analogy of high $T_{C}$ cuprates, $H_{C1}$ in Fig. \ref{Hc2} reflects mainly $H_{C1}^{H//ab}$ and both the $\lambda$ and $\kappa$ values should be considered as some averages of those along the $c$ axis and along the $ab$ plane as in the case of $\xi$. However, even taking into account this limitation, the phenomenological parameters obtained in the present study are quite unusual. It is strongly suggested that the present compound is an unconventional superconductor as expected by theoreticians\cite{Tanaka,Baskran}.

\begin{figure}
\includegraphics[width=7cm,keepaspectratio]{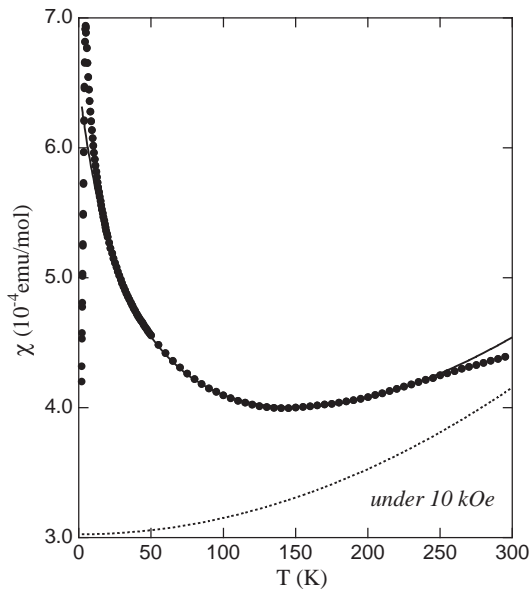}
\caption{
Temperature dependence of the magnetic susceptibility measured under 1 T. The solid line is the result when fitted by Eq. \ref{FitKai}. The dotted lines represents the Pauli paramagnetic terms of Eq. \ref{FitKai}. 
}
\label{kai}
\end{figure}
The magnetic susceptibility measured under 10 kOe is shown in Fig. \ref{kai}. The susceptibility decreases first with decreasing temperature and then increases between $T_{C}$ and 130 K. Assuming this upturn is due to a magnetic impurity and/or crystal defects, we fit the data between 20 K and 250 K by the equation:
\begin{equation}
\chi=\chi _{0}+AT^{2}+\frac{C}{T-\theta},
\label{FitKai}
\end{equation}
where the first and the second terms are due to Pauli paramagnetism. The parameters obtained are $\chi _{0}= 3.02\times 10^{-4}$ emu/mol, $A=1.25\times 10^{9}$ emu/K$^{2}$mol, $C=1.31\times 10^{-2}$ emu$\cdot$K/mol, and $\theta=-37.6$ K. The solid line in Fig. \ref{kai} represents Eq. \ref{FitKai} and seems to reproduce the data well. 
\begin{figure}
\includegraphics[width=7cm,keepaspectratio]{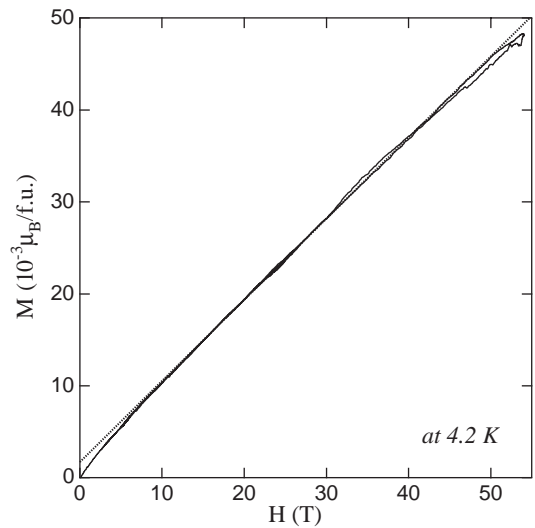}
\caption{
Magnetization curve at 4.2 K. The dotted line is the result when fitted by a linear function.
}
\label{HFMH}
\end{figure}
On the other hand, as seen in Fig. \ref{HFMH}, high-field magnetization increases linearly with increasing field above 15 T. Thus, if the Curie-Weiss-like upturn is caused by the magnetic impurity, its magnetization must be saturated above 15 T. In other words, the slope of the magnetization above 15 T should be equal to the magnetic susceptibility corresponding to the Pauli paramagnetism terms. However, the value of $4.9\times 10^{-4}$ emu/mol, which was calculated from the slope of the magnetization between 15 T and 30 T, is different from $3.02\times 10^{-4}$ ($=\chi _{0}+A\times 4.2^{2}$) emu/mol, obviously beyond experimental uncertainty. This result suggests that at least a part of the upturn reflects the intrinsic nature of the compound. The magnetic susceptibility of Na$_{0.5}$CoO$_{2}$ also increases with decreasing temperature owing to a kind of spin fluctuation\cite{NaCo2O4}. As an example showing such an enhancement of susceptibility, the La-doped Sr$_{2}$RuO$_{4}$ has been reported\cite{Sr214La} and its mother oxide is a spin triplet superconductor\cite{Sr214}. As well as the unconventional superconducting parameters, this enhancement may have relevance to the mechanism of the Cooper pair formation\cite{Tanaka}.

In summary, from $M$-$T$ and $M$-$H$ curves, superconducting transition temperature, as well as upper- and lower-critical fields are estimated to be $T_{C}=4.6$ K, $H_{C2}=61.0$ T and $H_{C1}=28.1$ Oe, for the novel Co oxide superconductor of Na$_{0.35}$CoO$_{2}$$\cdot$1.3H$_{2}$O. The coherent length, penetration depth and GL parameter are calculated to be $\xi=2.32$ nm, $\lambda=5.68\times 10^{3}$ \AA\ and $\kappa=244$. These phenomenological parameters are quite unusual and strongly suggest that the superconductivity is unconventional. The normal-state magnetic susceptibility is also presented for the first time. It shows an upturn below 130 K and at least a part of the upturn is suggested to be inherent. This normal-state magnetic behavior may have relevance to the mechanism of the Cooper pair formation.

\begin{acknowledgments}
We would like to thank A. Tanaka, T. Hikihara, X. Hu, M. Arai, and K. Kobayashi at NIMS and K. Yoshimura and T. Waki at Kyoto University for their useful advice and discussion. One of the authors (H.S.) was supported by the Japan Society for the Promotion of Science.
\end{acknowledgments}


\end{document}